\documentclass[12pt]{iopart}

\usepackage{iopams}  
\usepackage{graphicx}	
\usepackage{amssymb}	
\usepackage{mathrsfs}
\usepackage{soul,xcolor}

\newcommand*{\eqref}[1]{(\ref{#1})}

\begin{document}

\title[Response of nuclear-dissociating shocks to vorticity perturbations]{Response of nuclear-dissociating shocks to vorticity perturbations}

\author{C\'esar Huete$^{1}$ and Ernazar Abdikamalov$^{2}$}

\address{$^{1}$ Grupo de Mec\'anica de Fluidos, Universidad Carlos III, Av. Universidad 30, 28911, Legan\'es, SP\\
$^{2}$ Department of Physics, School of Science and Technology, Nazarbayev University, Astana 010000, KA \\

}
\ead{chuete@ing.uc3m.es}
\vspace{10pt}
\begin{indented}
\item[]August 2018
\end{indented}

\begin{abstract}
In the context of core-collapse supernova explosions (CCSNe), the interaction of standing accretion shocks with upstream vorticity perturbations is investigated by linear theory analysis. The endothermic effect associated to the nuclear dissociation, which takes place right behind the shock wave, affects the amplitude of the perturbations amplified/generated across the front. For upstream disturbances whose characteristic size is much larger than the post-shock dissociation-layer thickness, the effect of nuclear dissociation can be reduced to that of considering the global endothermic effect that scales with the inflow energy flux. \textcolor{black}{The present study focuses on perturbation fields that are not isotropic, which mimic the perturbations in collapsing convective shells of massive stars. The linear interaction of the shock with bidimensional mono-frequency vorticity perturbations is theoretically addressed, with the limit of highly-stretched vortices being analyzed in detail. The exact spatial distribution of the rotational and acoustic perturbations generated in the post-shock flow are provided along with the transient evolution of the shock front. It is found that nuclear dissociation contributes to stabilize the shock oscillations, but increases the amplitude of the density perturbations downstream.} An extension of this work that addresses the interaction with tridimensional isotropic turbulent flows can be found in reference Huete, C., {\it et al.} 2018, MNRAS, 475, 3305–3323, which analyzes the effect of the post-shock flow on the critical conditions that ultimately trigger explosion.
\end{abstract}

\vspace{2pc}

\submitto{\PS}

 \maketitle

%

\section{Introduction}

A distinguishing feature of massive stars is that the nuclear fusion occurring in their cores continues even after the hydrogen fuel is exhausted. The high temperatures induced by the strong gravity allow heavier elements such as helium and carbon to fuse sequentially. The energy released due to the mass defect between products and reactants keep the star stable, extending the lifetime of the star. Nonetheless, the characteristic time of these advanced fusion stages decreases very rapidly with the nuclear mass, with heavier nuclei burning on a timescale orders of magnitude shorter than the hydrogen sequence, which counts in million of years \cite{whw02}. In stars with initial masses $\sim (8-100)M_\odot$, the sequential nuclear fusion lasts until the formation of iron nuclei, a point beyond which nuclear fusion is no longer exothermic. As a result, the iron nuclei accumulates in the center, forming a core supported by the pressure of degenerate electrons. When the core reaches its maximum mass of $\simeq 1.4 M_\odot$, pressure begins to decay and the hydrodynamical stability breaks down, triggering a collapse of the iron core (e.g., \cite{Janka12} for a recent review).

\begin{figure}
\centering
\includegraphics[width=0.75\textwidth]{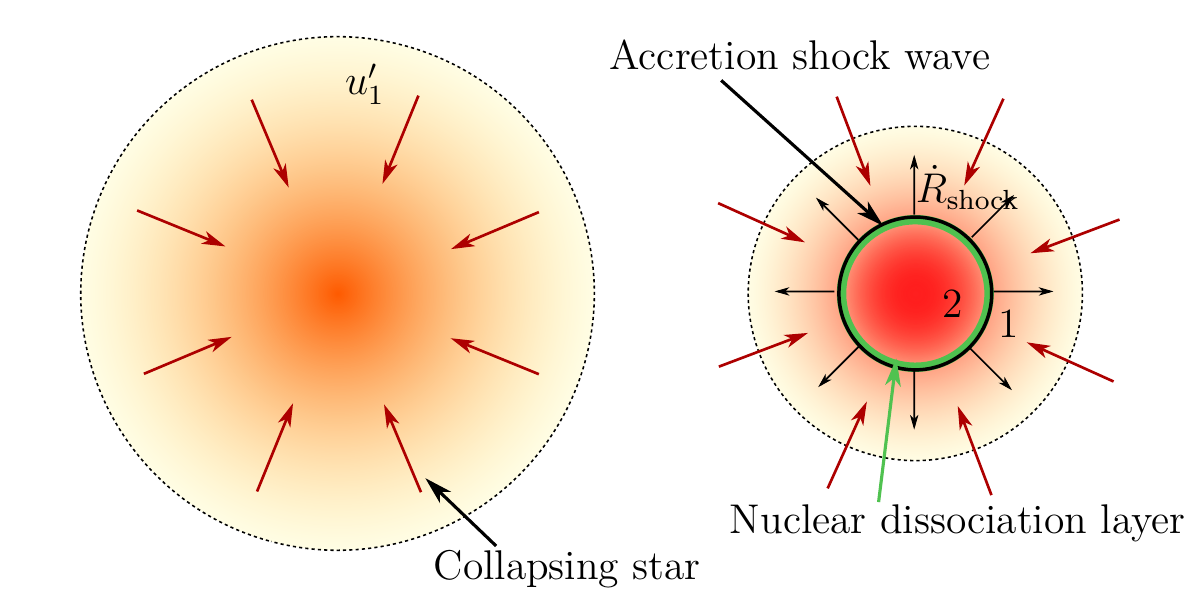}
\caption{Scheme of the shock formation when gravity attraction overcomes pressure forces. The supersonic in-falling matter is stoped and highly compressed by the shock wave. The shock induces heavy-nuclei breaking in the compressed matter, which is ultimately translated into an endothermic effect induced by the shock structure.}
\label{fig:scheme}
\end{figure}

The core collapse accelerates until the central density becomes as high as nuclear density ($ \sim 2 \times 10^{14} \, \mathrm{g/cm^3} $), a point where nuclear matter stiffens. This abruptly halts the collapse of the inner iron core, leading to the formation of a shock wave at the boundary of the inner core. The shock has to expel the stellar envelope and thus power core-collapse supernova (CCSN) explosion, leaving behind a stable neutron star (NS). The propagation of the shock, however, does not progress smoothly. The inherent pressure and temperature rise across the shock produces heavy-nuclei breaking as it propagates outwards, with associated energy consumption. In addition, the hot material behind the shock cools rapidly due to copious neutrino emission. As a result, the shock quickly loses its energy and turns in a stalled accretion shock within milliseconds after formation (see the sketch in Figure~\ref{fig:scheme}). Despite the decades of effort, the details of how to revive the shock and power CCSN explosion remain unclear (e.g., \cite{Burrows13,Janka16} for recent reviews). 


Along with the non-trivial, yet conventional, gas-dynamics effects, CCSNe is a very rich problem that comprises many different phenomena. The newly-born NS cools and contracts, releasing $\sim 10^{53}\,\mathrm{erg}$ potential binding energy as neutrino radiation. The shock-compressed matter is sufficiently opaque to absorb a small fraction of these neutrinos (e.g., \cite{Lentz12,Kotake18}). The deposited neutrino energy plays a key role in powering the $\sim 10^{51}\,\mathrm{erg}$ explosion\footnote{Core-collapse supernovae with explosion energies as high as $\sim 10^{52}\,\mathrm{erg}$ have also been observed. Also known as hypernovae, such explosions are relatively rare and are believed to be powered by the rotational kinetic energy of rapidly rotating protoneutron stars (e.g., \cite{Burrows07,Moesta14}).}. Neutrino heating leads to negative gradient of entropy, driving vigorous {\it neutrino-driven turbulent convection} in the postshock region (\cite{Mueller16,Radice18} for recent reviews), exerting additional pressure behind the shock \cite{Murphy13}. In addition, due to the {\it standing accretion shock instability} (SASI) \cite{Blondin03, Foglizzo06, Foglizzo07}, the shock undergoes large-scale non-radial oscillations \cite{Fernandez15,Hanke13}, which expands the size of the region subject to net neutrino heating, resulting in higher heating efficiencies. 

\textcolor{black}{Recently, it was shown that the convective instabilities that develop in the innermost nuclear-burning shells of massive stars can have an important impact on the explosion \cite{Couch13, Couch15a, Couch15, Mueller15, Mueller16a, Nagakura18}. The convective motion in oxygen and silicon shells may persist even during iron core collapse. As the core collapses, these shells contract and, due to the conservation of angular momentum, the velocities amplify as $\propto r^{-1}$, resulting in an increase by a factor of several (e.g., \cite{Abdikamalov16, Huete2018}). Upon reaching the shock, these perturbations interact with the supernova shock and generate additional turbulence in the post-shock region. This augments the turbulent pressure behind the shock (e.g., \cite{Mueller15}), resulting in more favorable conditions for producing explosion.}


The details of how these perturbations interact with the shock affects the properties of the resulting supernova explosion. Using the {\it linear interaction analysis} (LIA), Abdikamalov et al. \cite{Abdikamalov16,Abdikamalov18} studied the effect of acoustic, entropy, and vorticity perturbations, which are the three components of a generic weak hydrodynamic turbulent flow \cite{Kovasznay53}. Huete et al. \cite{Huete2018} improved their models by taking into account the perturbation of nuclear dissociation energy at the shock.  They employed long-time asymptotic expressions to compute the turbulent amplification ratios across the shock front for incoming vorticity waves. The impact of these modes on the explosion condition can be assessed using the notion of the critical (\textit{i.e.}, the minimum) neutrino luminosity necessary for driving the explosion \cite{Mueller15}. The effect of the entropic-buoyant turbulent perturbations generated by incident vorticity waves was found to reduce the critical luminosity by $\sim$17--24 per cent, which approximately agrees with the results of three-dimensional simulations of CCSNe \cite{Mueller17}. The present study is an extension of \cite{Huete2018} and it focuses on the linear interaction of the shock with bidimensional single-mode vorticity perturbations. Employing the mathematical formalism used in describing Richtmyer-Meshkov-type flows \cite{Wouchuk2001,Wouchuk2001b,Cobos2014} and perturbed non-reactive and reactive shocks \cite{Jackson1990,Velikovich07,Wouchuk2009,Huete2010,Huete2013,Huete2017}, the exact spatial distribution of the rotational and acoustic perturbations generated in the post-shock flow are provided along with the transient evolution of the shock front towards the permanent oscillatory mode, which, akin to non-ideal gases \cite{Bates2004,Bates2007}, may change the character of the decay when nuclear dissociation is sufficiently high. The effects of the nuclear dissociation energy, the shock strength, and the characteristic frequency are analyzed in the perturbed flow. \textcolor{black}{Finally, due to accelerated pace of stellar collapse, the inner parts of the star collapses faster than the outer parts (e,g, \cite{Bruenn85}). As a result, the convective vortices undergo substantial stretching during this phase. Such highly elongated vortices are also considered in this work.}

The paper is structured as follows: the problem formulation is shown in Section \ref{S2}, where the base-flow equations are presented and the linear-perturbation version are provided. The resulting Euler equations are integrated in Section \ref{S3}, where the transient and the long-time response of the shock front is computed. The complete distribution of the perturbations in the post-shock flow is also shown. \textcolor{black}{The transient evolution of the shock front in the limit of highly elongated vortices is addressed in Section \ref{S4}}. A summary of the results is provided in Section \ref{S5}.

\section{Problem Formulation}\label{S2}

\subsection{Base-flow equations}
Consider a shock wave located at radius $r=R_{\rm shock}(t)$ and assume that the shock thickness $l$ is much smaller than the shock radius ($l \ll R_{\rm shock}$). In this thin-shock limit, one can relate hydrodynamic quantities in the postshock region with those in the preshock region using the conservation equations for the mass, momentum and energy:
\numparts
\begin{eqnarray}
&\rho_1 \left(u_1'+\dot{R}_{{\rm shock}}\right) = \rho_2 \left(u_2'+\dot{R}_{{\rm shock}}\right) ,\label{mass0}\\
&p_1 + \rho_1 \left(u_1'+\dot{R}_{{\rm shock}}\right)^2 = p_2 + \rho_2 \left(u_2'+\dot{R}_{{\rm shock}}\right)^2,\label{momentum0}\\
&e_1 +\frac{p_1}{\rho_1} +\frac{1}{2} \left(u_1'+\dot{R}_{{\rm shock}}\right)^2 = e_2 +\frac{p_2}{\rho_2} + \frac{1}{2} \left(u_2'+\dot{R}_{{\rm shock}}\right)^2.\label{energy0}%
\end{eqnarray}
\endnumparts
Here, the flow ahead of the shock ($r>R_{\rm shock}$) is denoted with subscript 1, while the flow behind ($r<R_{\rm shock}$) is marked with subscript 2. The variable $u'$ is the bulk velocity measured in a reference frame at rest with respect to the center of the star, while variables $\rho$, $p$ and $e$ represent the density, pressure, and internal energy. We model stellar matter as a perfect gas with the polytropic index $\gamma=4/3$ on both sides of the shock. Since we consider scenario of in-falling flow and expanding shock, it is natural to define velocities positive, $u_1'>0$, $\dot{R}_{{\rm shock}}>0$, for the gas moving inwards and for the shock expanding outwards. 

In the presence of nuclear dissociation in a thin layer behind the shock front, the change of the internal energy across the shock is
\begin{equation}
e_1-e_2 =\frac{1}{\gamma-1}\frac{p_1}{\rho_1}-\frac{1}{\gamma-1}\frac{p_2}{\rho_2} + \Delta e_\mathrm{dis},
\end{equation}
where $\Delta e_\mathrm{dis}$ is the specific energy employed in the nuclear dissociation process. For stalled shock in CCSNe, $\Delta e_\mathrm{dis}$ can be parametrized as $\Delta e_\mathrm{dis} =\varepsilon \upsilon_\mathrm{FF}^2/2$, where $\upsilon_\mathrm{FF}$ free-fall speed and $\epsilon$ is a dimensionless parameter \cite{Fernandez2009a,Fernandez2009b}. In this scenario, $\epsilon$ scales as $\sim 0.67 {\cal M}_{1.3}^{-1}\left(R_{\rm shock}/150\ {\rm km}\right)$, which results in $\varepsilon$ typically ranging between $0.2$ and $0.5$ \cite{Huete2018}. For flows with vanishing Bernoulli parameter above the shock, one can express $\Delta e_\mathrm{dis}$ in terms of the preshock Mach number $M_1 =u_1/a_1$ (see \cite{Huete2018} for the details of the derivation),
\begin{equation}
\Delta e_\mathrm{dis} = \varepsilon \frac{a_1^2}{\gamma-1}\left(1+\frac{\gamma-1}{2}M_1^2\right),
\label{eq:varepsilon}
\end{equation}
where $a_1=(\gamma_1 p_1/\rho_1)^{1/2}$ is the sound speed in the preshock region and $u_1=u_1'+\dot{R}_{{\rm shock}}$ is the preshock speed in the reference frame at rest with respect to the shock.

The fluid properties behind the shock can be conveniently expressed as functions of $\varepsilon$ and preshock Mach number $M_1$:
\begin{equation}
C_2 = \frac{\rho_2}{\rho_1} = \frac{u_1}{u_2} = \frac{\left( \gamma + 1\right) M_1^2}{\left( \gamma - \kappa  \right) M_1^2 + 1},
\label{R}
\end{equation}
\begin{equation}
P_2 =\frac{p_2}{\rho_1 u_1^2}=\frac{\gamma M_1^2(1+\kappa) +1}{\gamma(\gamma + 1) M_1^2},
\label{P}
\end{equation}
\begin{equation}
M_2 = \frac{u_2}{a_2} = \left(\gamma C_2 P_2\right)^{-1/2} =  \left[\frac{\left( \gamma - \kappa  \right) M_1^2 + 1}{\gamma M_1^2(1+\kappa) +1}\right]^{1/2},
\label{M2}
\end{equation}
where $M_2$ is the mean Mach number in the post-shock region and $u_2=u_2'+\dot{R}_{{\rm shock}}$ is the speed of the postshock flow in the reference frame at rest with respect to the shock. The function 
\begin{equation}
\kappa=\left[(1-M_1^{-2})^2+ \varepsilon(\gamma+1) \left(\gamma-1+2 M_1^{-2}\right)\right]^{1/2}
\label{kappa}
\end{equation}
contains the effect of nuclear dissociation. In the limit of vanishing nuclear dissociation, $1-\kappa \sim M_1^{-2}$, expressions \eqref{R}-\eqref{M2} reduce to the classical Rankine-Hugoniot relations.

\begin{figure}
\centering
\includegraphics[width=0.75\textwidth]{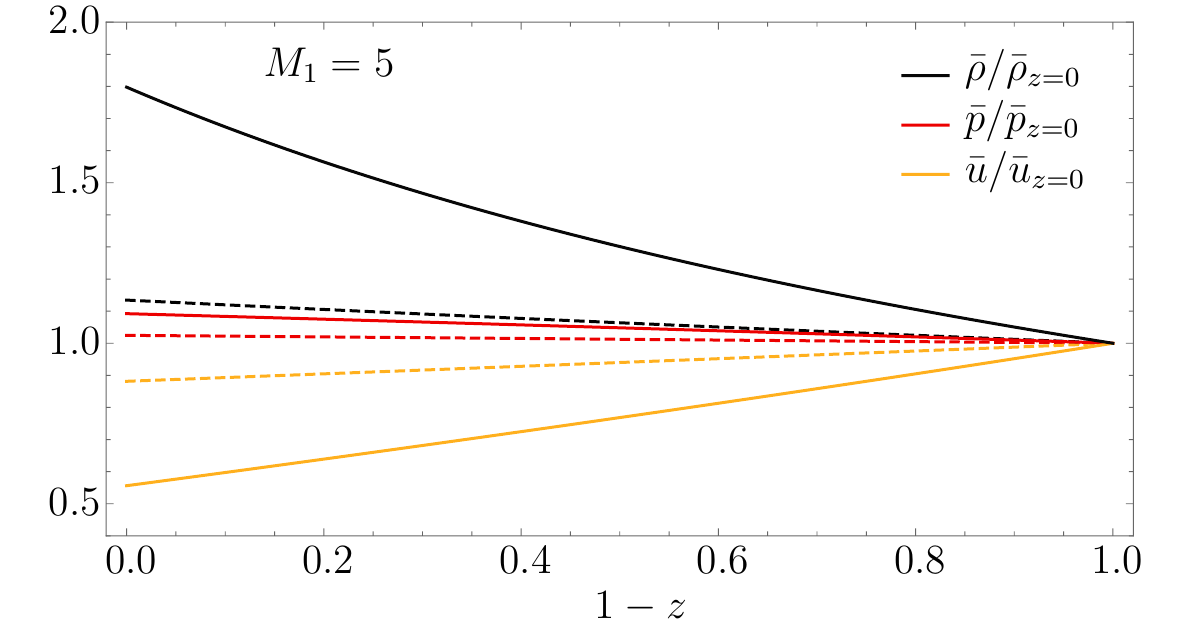}
\caption{Flow properties inside the dissociation layer as a function of the nuclear-breaking progress variable $z$. The conditions of the computations are $M_1=5$ and $\varepsilon = 0.1$ (dashed lines) and $\varepsilon=0.4$ (solid lines).}
\label{fig:profiles}
\end{figure} 

The progress of the nuclear dissociation taking place behind the shock can be quantified in terms of variable $z$, such that $z=0$ refers to values right behind the adiabatic shock and $z=1$ refers to the corresponding properties once nuclear dissociation has been accomplished. Normalized with the flow properties right behind the shock, which are directly obtained from~\eqref{R}-\eqref{kappa} with $\varepsilon=0$, the inner properties are computed as a function of $z$ in Figure \ref{fig:profiles}. The effect of nuclear dissociation is to increase the density ratio with the corresponding velocity decrease, and also to slightly increase the downstream pressure. The rate of the heavy-nuclei breaking $\dot{z}$ would provide the characteristic reaction time, whose combination with the characteristic velocity $a_1$ can be used to scale dissociation layer thickness $\ell\sim a_1/\dot{z}$.

\subsection{Linear perturbation analysis}

\begin{figure}
\centering
\includegraphics[width=0.85\textwidth]{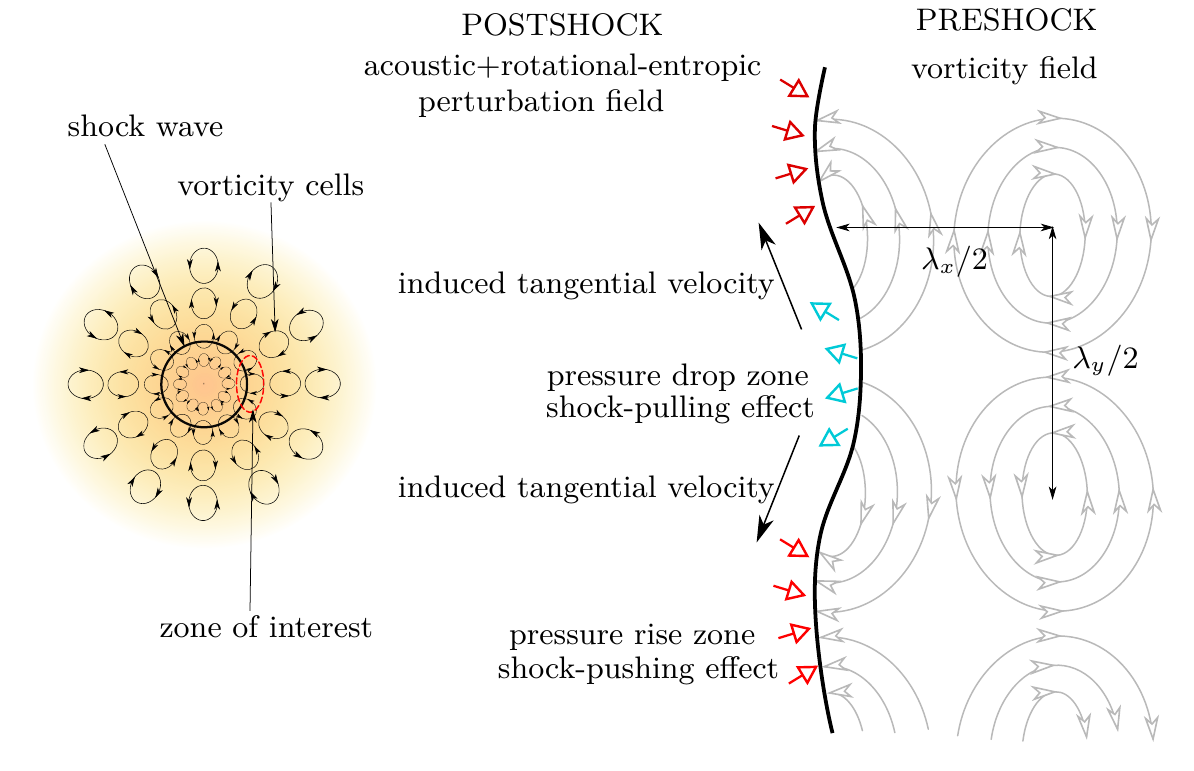}
\caption{Scheme of the interaction of the shock front with the mono-frequency vorticity field upstream in the planar-shock limit $R_{\rm shock}\gg\lambda_y\sim \lambda_x$. Induced transverse velocity behind the shock generates a counter-pressure effect to balance the shock deformation.}
\label{fig:shock_vor}
\end{figure}

The upstream flow is seldom uniform. The in-falling matter is immersed in a strong gravity field that triggers convective instabilities. Vortex cells are then formed upstream and they perturb the shock front. The postshock flow is correspondingly perturbed. In order to study how the shock wave reacts to perturbations, a canonical case is selected to be studied: the interaction of harmonic vorticity perturbations with the shock wave in the planar fast-reaction limit $R_{\rm shock}\gg\lambda_y\sim \lambda_x\gg\ell$. Such an interaction is sketched in Figure \ref{fig:shock_vor}, where the disturbed shock induces pressure, density and velocity changes in the flow downstream. 

The amplitude of the velocity perturbations is assumed to be much smaller than the background flow properties. The small dimensionless amplitude factor $\hat{u}_1\sim \left(u_1-\langle u_1 \rangle\right)/\langle u_1 \rangle  \ll 1$, is used to scale the preshock and postshock perturbation variables. The incident shear wave in the frame $(x_1,y_1)$ comoving with the in-falling fluid particles is expressed as a divergence-free velocity field
\numparts
\begin{eqnarray}
&\bar{u}_1 \left( x_1, y_1 \right)= \frac{u_1-\langle u_1 \rangle}{\hat{u}_1 \langle a_2 \rangle}
= \cos\left(k_x x_1 \right)\cos\left(k_y y_1 \right), \label{u1v10} \\
&\bar{v}_1 \left( x_1, y_1 \right)= \frac{v_1-\langle v_1 \rangle}{\hat{u}_1 \langle a_2 \rangle}
=  \frac{k_x}{k_y} \sin\left(k_x x_1 \right)\sin\left(k_y y_1 \right),
\label{u1v1}%
\end{eqnarray}
\endnumparts
where $\bar{u}_1$ and $\bar{v}_1$ are the order-unity stream-wise and transverse velocity perturbation components. The angle brackets represent the time-averaged value of the flow variables. The dimensionless vorticity function, associated to the rotational velocity perturbation (\ref{u1v10})-(\ref{u1v1}), is 
\begin{equation}
\bar{\omega}_1 \left(x_1, y_1 \right) =\frac{\partial \bar{v}_1}{\partial (k_y x_1)}- \frac{\partial \bar{u}_1}{\partial (k_y y_1)}= \left(1+\frac{k_x^2}{k_y^2}\right)  \cos\left(k_x x_1 \right)\sin\left(k_y y_1 \right),
\label{omega1}
\end{equation}
where $\vec{k}=(k_x,k_y)$ is the perturbation wavenumber in the pre-shock region, which can be expressed in terms of the wavelengths $k_x=2\pi /\lambda_x$ and $k_y=2\pi /\lambda_x$ sketched in Figure~\ref{fig:shock_vor}, or in terms of the incident shear angle $\theta = \tan^{-1}(k_y/k_x)$. For the analysis of the postshock flow, it is most natural to use a reference frame comoving with the postshock flow. For this reason, hereafter, the dimensionless coordinates $x = k_y x_2$ and $y = k_y y_2$ and the dimensionless time $\tau = a_2 k_y t $ are used to describe the solution in the postshock region.

As a result of the interaction, the shock ripples and the fluid downstream is correspondingly altered with acoustic and entropic-vorticity waves. The former travels at the speed of sound downstream $a_2$ and the latter moves with the fluid particles. In the postshock region, the dimensionless pressure, density and velocity perturbations are defined as order-unity functions
\begin{equation}
\bar{p} = \frac{p-\langle p_2 \rangle}{\hat{u}_1\gamma \langle p_2 \rangle} ,\quad 
\bar{\rho}= \frac{\rho-\langle \rho_2 \rangle}{\hat{u}_1\langle \rho_2 \rangle}, \quad
\bar{u}=  \frac{u-\langle u_2 \rangle}{\hat{u}_1\langle a_2 \rangle} ,\quad 
\bar{v}= \frac{v-\langle v_2 \rangle}{\hat{u}_1\langle a_2 \rangle}.
\label{perturb}
\end{equation}
The adiabatic Euler equations governing the postshock flow are written in terms of these variables. Since $\bar{p}$ and $\bar{v}$ are proportional to $\cos(y)$ and $\sin(y)$, the conservation equations for mass, $x$-momentum, $y$-momentum, and energy
\begin{equation}
\frac{\partial \bar{\rho}}{\partial \tau} +\frac{\partial \bar{u}}{\partial x}+\bar{v}=0 ,\quad 
\frac{\partial \bar{u}}{\partial \tau} +\frac{\partial \bar{p}}{\partial x}=0 ,\quad 
\frac{\partial \bar{v}}{\partial \tau} -\bar{p}=0 ,\quad 
\frac{\partial \bar{p}}{\partial \tau} = \frac{\partial \bar{\rho}}{\partial \tau},
\label{eqperturb}
\end{equation}
can be combined to yield an equation for pressure perturbation $\bar{p}$:
\begin{equation}
\frac{\partial^2 \bar{p}}{\partial \tau^2}= \frac{\partial^2 \bar{p}}{\partial x^2}-\bar{p}.
\label{sonicwave}
\end{equation}
This is a periodically-symmetric two-dimensional wave equation, the solution of which yields the perturbation field in the postshock region. 

The problem thus reduces to solving the wave equation~\eqref{sonicwave} as a function of $x$ and $\tau$ for $\tau>0$. The spatial domain is bounded by the leading (first) reflected acoustic wave propagating backwards, $x= -\tau$, and the shock wave front traveling upwards, $x= M_2\tau$. One of the boundary conditions stems from the assumption of isolated shock, according to which no reflected sonic waves can reach the shock from behind. This is consistent with the large-radius approximation, $R_{\rm shock} k_y \gg 1$, and the linear perturbation analysis that neglects the second-order interaction of these sonic waves with the entropic or rotational perturbations downstream. The upstream flow distorting the shock front is what ultimately determines the amplitude of the shock oscillations and the post-shock inhomogeneities. 

The boundary conditions at the shock are readily obtained from the linearized Rankine-Hugoniot relations \eqref{mass0}-\eqref{energy0} along with the conservation of momentum transverse to the shock front, namely
\numparts
\begin{eqnarray}
&\left(C_2-1\right)\dot{\xi}_s =C_2\bar{u}_s-M_2 C_2\bar{\rho}_s-\bar{u}_1 , \label{massRH}\frac{}{}\\
&\bar{p}_s = 2M_2\left(\bar{u}_s-\bar{u}_1 \right)-M_2^2\bar{\rho}_s, \label{xmomRH}\frac{}{}\\
&M_1^2 M_2^2\bar{\rho}_s = \Pi_s\bar{p}_s-\Delta_s \left(\dot{\xi}_s-\bar{u}_1\right), 
\label{eneRH}\frac{}{}\\
&\bar{v}_s = M_2 \left(C_2-1\right)\xi_s+\bar{v}_1,\label{tanRH}
\end{eqnarray}
\endnumparts
where $\dot{\xi}_s$ is the time derivative of the dimensionless shock front deformation. As sketched in Figure~\ref{fig:shock_vor}, the transverse velocity induced behind the shock, included in \eqref{tanRH}, generates a counter-pressure effect to balance the shock ripple. The characteristic shock reaction time would determine the shock evolution, which in turn is affected by the energy absorbed in the nuclear dissociation process. This effect is quantified in the energy equation~\eqref{eneRH} via functions
\begin{equation}
\Pi_s = \frac{M_1^2\left[1 + M_1^2\left(1-\kappa\right)\right]^2}{\left(M_1^2+1\right)^2-M_1^4\kappa^2} 
\label{Pis}
\end{equation}
and
\begin{equation}
\Delta_s = \varepsilon\frac{2 M_2 M_1^4\left(\gamma-1 \right) \left[1 + M_1^2\left(1-\kappa\right)\right]}{\left(M_1^2+1\right)^2-M_1^4\kappa^2},
\label{Deltas}
\end{equation}
that allow us to differentiate adiabatic shock waves from reacting shock waves such as shocks with nuclear dissociation or detonations. Functions \eqref{Pis} and \eqref{Deltas} are equivalent to those provided in \cite{Huete2018}, with the latter being here corrected by a factor $1/M_2^2 M_1^2$.

From \eqref{massRH}-\eqref{tanRH}, one can obtain one of the two relations for the shock boundary condition involving $\bar{\xi}_s$ and $\bar{p}_s$, while the other can be obtained using the material derivative of the longitudinal velocity perturbation behind the shock, namely
\numparts
\begin{eqnarray}
&\frac{d \xi_s}{d \tau} = \sigma_a \bar{p}_s+\cos\left(\frac{k_x}{k_y}C_2 M_2 \tau\right),\\
\left(\sigma_b +M_2 \right)\frac{\partial \bar{p}_s}{\partial \tau} &+\left.\left(\sigma_b M_2+ 1\right)\frac{\partial \bar{p}}{\partial x}\right|_s+M_2^2 \left(C_2- 1\right)\xi_s=\nonumber  \\
&+\frac{k_x}{k_y}M_2\left(C_2-1\right) \sin\left(\frac{k_x}{k_y}C_2 M_2 \tau\right),
\label{ps}
\label{xis}
\end{eqnarray}
\endnumparts
where
\begin{equation}
\sigma_a = \frac{C_2\left(M_1^2-\Pi_s\right)}{2 M_2 M_1^2  \left(C_2-1\right)+C_2\Delta_s},\quad \sigma_b = \frac{M_1^2+\Pi_s+\Delta_s \sigma_a}{2 M_2 M_1^2}
\label{sigmasaandb}
\end{equation}
are the factors accompanying the pressure perturbation.

The initial condition for the shock perturbations can be obtained from the requirement that the initial shock is planar,  \textit{i.e.}, $\bar{\xi}_s= \bar{v}_s = 0$. Consequently, the initial pressure and streamwise velocity perturbations obey condition $\bar{u}_s +\bar{p}_s=0$, thus yielding $\bar{p}_{s0}(\sigma_b+1)=1$, for the initial pressure perturbation right behind the shock.

\section{Results}\label{S3}
\subsection{Temporal evolution of the shock front}
In order to study the transient response of the shock to upstream perturbation, the transformation 
\begin{equation}
x = r \sinh \chi, \qquad \tau = r \cosh \chi \label{rchi}
\end{equation}
is employed \cite{Wouchuk2009,Huete2017}. The initial condition, $\tau=0$, corresponds to the moment when the initial unperturbed shock first meets the incident vorticity perturbations. The $\chi=$const condition represents a planar surface moving in the postshock gas along the $x$ direction, from the weak discontinuity at $x=0$ $(\chi=0)$ to the reacting shock front at $x=M_2 \tau$ $(\tanh \chi_s=M_2)$. In terms of these variables, equation~\eqref{sonicwave} for sound waves reads
\begin{equation}
r\frac{\partial^2 \bar{p}}{\partial r^2} + \frac{\partial \bar{p}}{\partial r} + r \bar{p} = \frac{1}{r}\frac{\partial^2 \bar{p}}{\partial \chi^2}.
\label{sonicrchi}
\end{equation}
The boundary conditions at the shock front reduces to 
\begin{equation}
\frac{d \xi_s(r)}{d r} = \frac{\sigma_a}{\sqrt{1-M_2^2}} \bar{p}_s(r)+ \frac{1}{\sqrt{1-M_2^2}}\cos\left(\zeta r\right)
\label{xisrp}
\end{equation}
and
\begin{equation}
\left.\frac{1}{r}\frac{\partial \bar{p}_s}{\partial \chi}\right|_s = -\sigma_b \frac{\partial \bar{p}_s(r)}{\partial r}-\frac{M_2^2 \left(C_2- 1\right)}{\sqrt{1-M_2^2}}\xi_s(r) +\zeta \frac{C_2-1}{C_2}\sin\left(\zeta r\right),
\label{psr}
\end{equation}
where
\begin{equation}
\zeta = \frac{k_x}{k_y}\frac{M_2 C_2}{\sqrt{1-M_2^2}}  = \frac{\omega_s}{\sqrt{1-M_2^2}}=\frac{1}{\tan\theta}\frac{M_2 C_2}{\sqrt{1-M_2^2}}
\label{zeta}
\end{equation}
is the characteristic shock oscillation frequency induced by the incident shear wave.

The Laplace transform is conveniently employed to reduce the above system of partial differential equations to an algebraic system. That is, the integral
\begin{equation}
\mathscr{F}(s,\chi) = \int_0^{\infty}f(r,\chi)e^{-sr}\rm{d} r
\label{Laplace}
\end{equation}
applied to the functions defining the shock boundary conditions yields an algebraic system of function $s$. The Laplace transform of $\left.\frac{1}{r}\frac{\partial \bar{p}_s}{\partial \chi}\right|_s $ can be computed using the isolated boundary condition, namely $\sqrt{s^2+1}\mathscr{P}_s - \bar{p}_{s0}$. From this, one can obtain the Laplace transform of the pressure perturbation at the shock:
\begin{equation}
\mathscr{P}_s(s) =\frac{s \left(1+\sigma_b\right)\bar{p}_{s0}}{s\sqrt{s^2+1}+\sigma_b s^2 + \sigma_c}+ \frac{s \sigma}{\left(s\sqrt{s^2+1}+\sigma_b s^2 + \sigma_c\right)\left(s^2+\zeta^2\right)},
    \label{PsLaplace}
\end{equation}
where
\begin{equation}
\sigma = \frac{C_2-1}{C_2}\left( \zeta^2 -\frac{M_1^2}{M_1^2-1}\right)
\label{sigma}
\end{equation} 
is the factor accounting for the periodic excitation amplitude, and 
\begin{equation}
\sigma_c = \frac{M_2^2\left(C_2-1\right)}{1-M_2^2}\sigma_a.
\label{Cepsilon}
\end{equation}
The solution of equation ~\eqref{sonicrchi} for the pressure field can be expressed as a combination of the Bessel functions \cite{Zaidel1960}, as shown in \cite{Huete2018} in this particular context. It is however illustrative to use the inverse of the Laplace transform to the function \eqref{PsLaplace}, which yields
\begin{equation}
\bar{p}_s(r) = -\frac{2}{\pi}\int_0^1 \cos(z r)f(z) {\rm d}z+\frac{2\sigma}{\pi}\int_0^1 \frac{\cos(z r)-\cos(\zeta r)}{\zeta^2-z^2}f(z) {\rm d}z 
\label{pr}
\end{equation}
as the temporal evolution of the shock pressure perturbations, with 
\begin{equation}
f(z) = \frac{z^2\sqrt{1-z^2}}{z^2(1-z^2)+\left(\sigma_b z^2-\sigma_c\right)^2} 
\label{fz1}
\end{equation}
being the auxiliary function. The corresponding $\bar{p}_s(\tau)$ is readily given by the variable change $r=\sqrt{1-M_2^2}\ \tau$.

\begin{figure}
\centering
\includegraphics[width=0.75\textwidth]{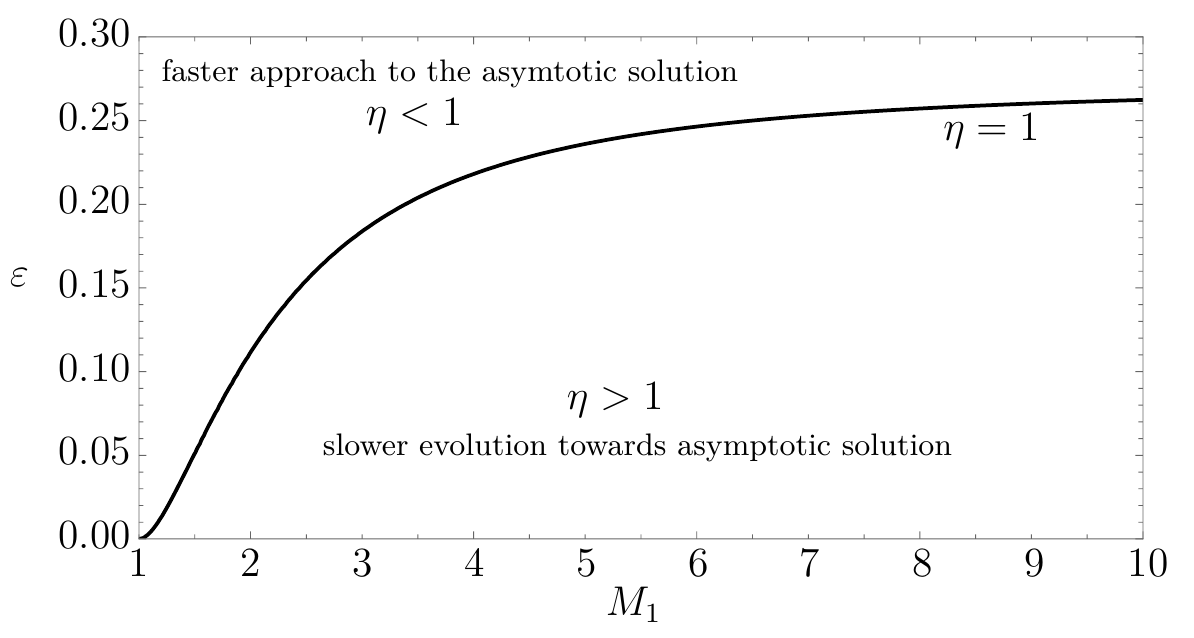}
\caption{Iso-curve $\eta=1$ as a function of the shock strength $M_1$ and the endothermic parameter $\varepsilon$. In the upper region (strong endothermicity), the shock dynamics towards asymptotic solution is shorter than that in the lower region (weak endothermicity).}
\label{fig:Bates}
\end{figure} 

Self-induced stable oscillations are found to depend on the slope of the Rankine-Hugoniot curve \cite{Dyakov1954,Kontorovich1957,Wouchuk2004,Bates2004,Bates2007,Clavin2012} and the corresponding spontaneous acoustic radiation is determined by the condition $\sigma_c>\sigma_b$. A parametrical study reveals that it would occur only for $\varepsilon<0$, \textit{i.e.}, when the net positive energy release increases with the shock intensity. In this case, as $\sigma_c<\sigma_b$ the shock will oscillate only with the excitement frequency coming from upstream perturbations, $\omega_s = C_2 M_2 k_x/k_y$, thereby yielding an asymptotic response qualitatively similar to the one found for adiabatic shock waves \cite{Wouchuk2009}. Nonetheless, the endothermic contribution may have a qualitative impact on the transient evolution towards the long-time dynamics. Although transient evolution always decays in time with $\tau^{-3/2}$ for $\varepsilon>0$, when the function 
\begin{equation}
\eta = \frac{\left(2 \sigma_b \sigma_c-1\right)^2}{4\left(\sigma_b^2-1\right)}
\label{eta}
\end{equation}
is lower than unity (or $\Lambda<0$ in \cite{Bates2004}), the initial degree of damping is significantly modified. That is, for $\eta<1$, corresponding highly endothermic shocks, the oscillations associated to the transient response are effectively shorten, while the contrary occurs for $\eta>1$, the latter case corresponding to the regular shock family of solutions. The delimiting curve $\eta=1$ is computed in Figure \ref{fig:Bates} for $\gamma=4/3$ as a function of the shock strength $M_1$ and the endothermic parameter $\varepsilon$. 
 
Irrespective of the transient behavior, the long-time reaction of the shock pressure to mono-frequency perturbations is
\begin{equation}
\bar{p}_{s}(\tau\gg 1) =  \left \{ \begin{array}{ll}
\mathcal{P}_{lr} \cos\left(\omega_s \tau \right) + \mathcal{P}_{li} \sin\left( \omega_s \tau \right) & ,\zeta \leq 1 \\
\mathcal{P}_{s} \cos\left(\omega_s \tau \right) & ,\zeta \geq 1
	\end{array} \right.
\label{pstau}
\end{equation}
where
\begin{equation}
\mathcal{P}_{lr} = \frac{-\sigma \left(\sigma_b\zeta^2-\sigma_c\right)}{\zeta^2 \left(1-\zeta^2\right)+\left(\sigma_b \zeta^2 - \sigma_c\right)^2}, \quad
\mathcal{P}_{li} = \frac{\sigma \zeta\sqrt{1-\zeta^2}}{\zeta^2 \left(1-\zeta^2\right)+\left(\sigma_b \zeta^2 - \sigma_c\right)^2}
\label{PlrPli}
\end{equation}
for $\zeta <1$, and 
\begin{equation}
\mathcal{P}_{s} = \frac{-\sigma }{\zeta\sqrt{1-\zeta^2}+\sigma_b \zeta^2 - \sigma_c}
\label{Ps}
\end{equation}
for $\zeta >1$. As in previous studies of the interaction of shocks with vorticity perturbations \cite{Ribner54,Lee97,Mahesh97,Wouchuk2009,Huete2018}, in the long wavelength regime ($\zeta <1$), the sonic disturbances immediately behind the shock consists of two orthogonal contributions $\mathcal{P}_{lr}$, and $\mathcal{P}_{li}$. In this regime, the sonic waves decay exponentially as they move away from the shock. In contrast to this, in the short wavelength regime ($\zeta>1$), the solution is represented by constant-amplitude sonic waves. The critical $\zeta=1$ value corresponds to the case when stable acoustic waves travel parallel to the shock surface in the shock reference frame.

Likewise, the integration of \eqref{xis} with respect to $\tau$ yields the temporal evolution of the shock oscillation amplitude:
\begin{eqnarray}
\xi_s(r) &= \frac{\sin(\zeta r)}{\zeta \sqrt{1-M_2^2}}-\frac{2\sigma_a}{\pi\sqrt{1-M_2^2}}\int_0^1 \sin(z r)\frac{f(z)}{z} {\rm d}z+\nonumber \\
&+\frac{2\sigma\sigma_a}{\pi\sqrt{1-M_2^2}}\int_0^1 \left(\frac{\sin(z r)}{z}-\frac{\sin(\zeta r)}{\zeta}\right)\frac{f(z)}{\zeta^2-z^2} {\rm d}z, 
\label{xisr}
\end{eqnarray}
with the variable $r=\sqrt{1-M_2^2}\ \tau$. The associated long-time function of the shock oscillations is
\begin{equation}
\xi_{s}(\tau\gg 1) =  \left \{ \begin{array}{ll}
\mathcal{J}_{lr} \sin\left(\omega_s \tau \right) + \mathcal{J}_{li} \cos\left( \omega_s \tau \right) & ,\zeta \leq 1 \\
\mathcal{J}_{s} \sin\left(\omega_s \tau \right) & ,\zeta \geq 1
	\end{array} \right.
\label{xisasym}
\end{equation}
where the coefficients $\mathcal{J}$ are obtained from the pressure fluctuations at the shock:
\begin{equation}
\mathcal{J}_{lr} = \frac{\sigma_a}{\omega_s}\mathcal{P}_{lr} + \frac{1}{\omega_s}, \quad 
\mathcal{J}_{li} = -\frac{\sigma_a}{\omega_s}\mathcal{P}_{li}, \quad\mathcal{J}_{s} = \frac{\sigma_a}{\omega_s}\mathcal{P}_{s} + \frac{1}{\omega_s}.
\label{Jlri}
\end{equation}

\begin{figure}
\centering
\includegraphics[width=\textwidth]{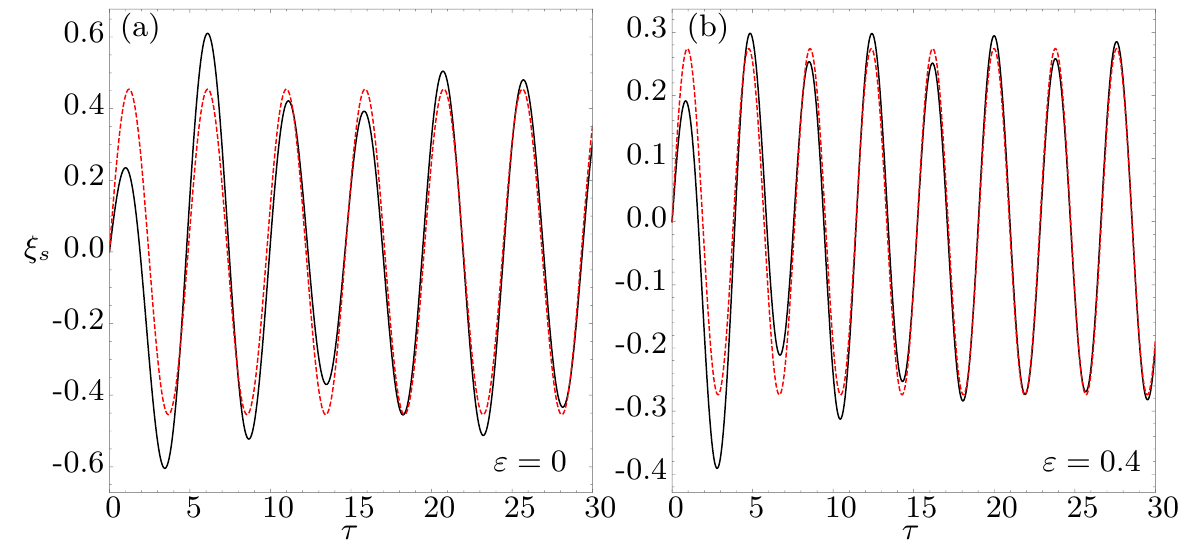}
\caption{Non-dimensional shock-ripple amplitude $\xi_s$ as a function of the dimensionless time $\tau$ for $M_1=5$ and $\theta = 60^\circ$, and for $\varepsilon=0.0$ (a) and $\varepsilon=0.4$ (b). Exact temporal evolution \eqref{xisr} in black-solid lines and asymptotic solution \eqref{xisasym} in red-dashed lines. The characteristic frequencies are $\zeta= 1.4$ and $\zeta= 1.72$, respectively.}
\label{fig:xi_asym}
\end{figure} 

A direct comparison of the long-time response, provided by \eqref{xisasym}, and the exact temporal evolution, given by \eqref{xisr}, is computed in Figure \ref{fig:xi_asym} as a function of $\tau$ for a shock strength $M_1=5$ with two different dissociation sensitivities $\varepsilon=0$ (a) and $\varepsilon=0.4$ (b). The figure is qualitatively similar to figure 5 in \cite{Huete2018}, where different input parameters had been selected in this occasion. The upstream shear-wave angle is $\theta = 60^\circ$, which yields dimensionless oscillation frequencies greater than unity in both cases: $\zeta= 1.4$ and $\zeta= 1.72$ for the adiabatic and endothermic case, respectively. The transient solution is found to achieve the asymptotic regime in a relatively short period of time, with the panel on the left doing it seemingly faster, in agreement with Figure \ref{fig:Bates}. As expected, the amplitude of the long-time oscillations is found to be smaller in the endothermic case. 

\begin{figure}
\centering
\includegraphics[width=\textwidth]{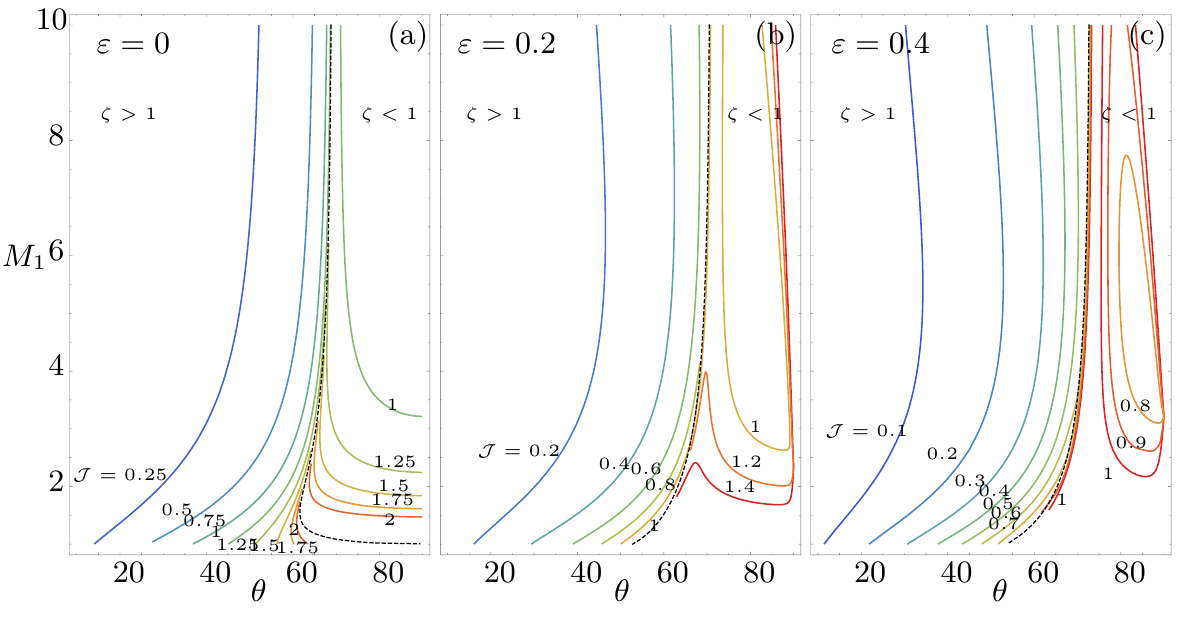}
\caption{Iso-curves of the asymptotic shock ripple amplitude~\eqref{Jlri} as a function of the shock strength $M_1$ and the incident wave angle $\theta$ for $\varepsilon=0$ (a), $\varepsilon=0.2$ (b), and $\varepsilon=0.4$ (c).}
\label{fig:J0}
\end{figure}

Figure \ref{fig:J0} shows the amplitude of the oscillations as a function of the shock strength $M_1$, shear-wave angle $\theta$, and for three different values of the dissociation degree $\varepsilon=0$ (a), $\varepsilon=0.2$ (b), and $\varepsilon=0.4$ (c). In agreement with Figure \ref{fig:xi_asym}, the amplitude is found to be generally smaller for endothermic shocks. The limit $\zeta=1$ is also computed in Figure \ref{fig:J0} as a function of $M_1$ and $\theta$ for different endothermic intensities. The zones on the left of these dashed curves correspond to pressure radiating conditions (high-frequency regime), and the zones on the right refer to non-radiating conditions (low-frequency regime), with the area of the latter being reduced with the increase of nuclear dissociation. 

\subsection{Post-shock flow variables}
The downstream flow perturbations are conveniently split decomposed those of acoustic type, which travel at the speed of sound relative to the fluid particles, and those of entropic-rotational nature, which move with the fluid particles \cite{Kovasznay53}. 

\begin{figure}
\centering
\includegraphics[width=\textwidth]{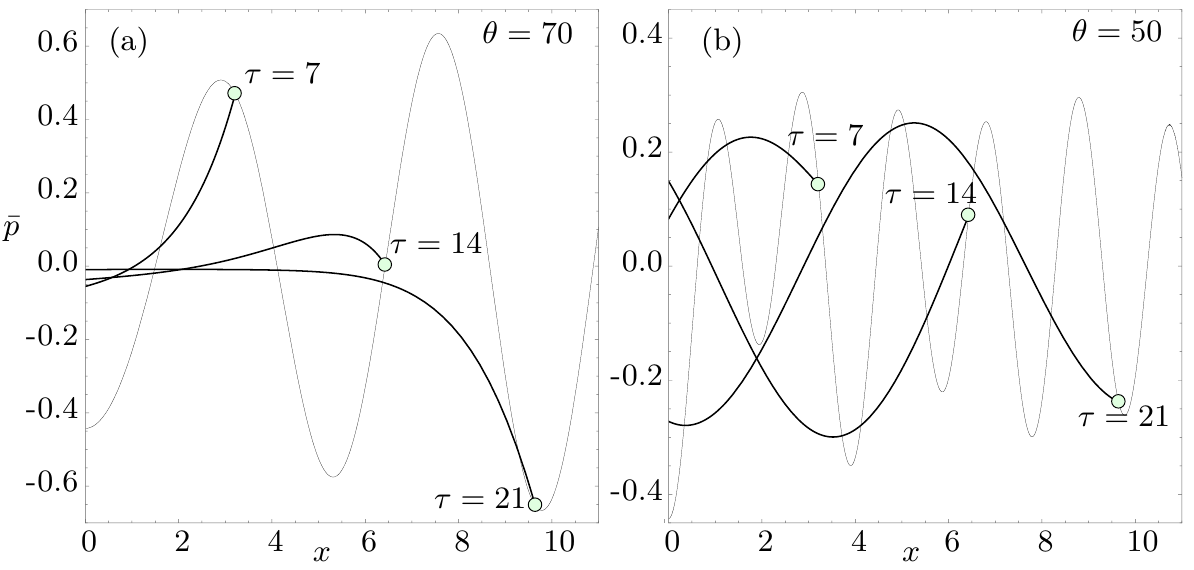}
\caption{The thick-black line shows the spatial distribution of the pressure field $\bar{p}$ for $M_1=2$, $\varepsilon=0.2$, and for $\tau=7$, $\tau=14$ and $\tau=21$. The thin line shows to the shock-pressure history. The left panel corresponds to non-radiating conditions $\theta =70^\circ$ while the panel on the refers to radiating conditions, $\theta =50^\circ$, in agreement with the panel in the middle of Figure \ref{fig:J0}.}
\label{fig:pressurefield}
\end{figure}

The exact temporal evolution of the pressure field downstream is readily obtained through Bessel functions, as derived in \cite{Huete2018}, or by direct integration of the wave equation provided that $\bar{p}(x=M_2\tau)=\bar{p}_s(\tau)$. The former strategy has been employed in Figure \ref{fig:pressurefield} to compute the pressure field in the domain $0\leq x \leq M_2\tau$. Computations reveal two well-distinguished regimes: acoustically radiating and non-radiating conditions. The acoustic radiation condition is then determined by $\omega_s>(1-M_2^2)^{1/2}$, a condition that depends on the upstream shear wave, since $\zeta=\left[0,\infty\right)$ depends on the relative properties of the perturbation field ahead of the shock. Small values of $\zeta$ represent the interaction with upstream vortices highly stretched in the streamwise direction $\lambda_x\gg\lambda_y$, while the opposite is true for $\zeta\gg 1$. In the latter low mode-number scenario ($\lambda_x\ll\lambda_y$), the problem reduces to the one-dimensional interaction of the shock with radial perturbation waves. 

The sonic waves traveling in the postshock region are functions of $(\omega_{a} \tau - k_{a} x)$, where the frequency $\omega_{a}$ and the wavenumber $k_{a}$ are obtained from the shock oscillation frequency $\omega_{s}= \omega_{a} - M_2 k_{a}$ and the adiabatic dispersion relation $\omega_{a}^2=k_{a}^2+1$:
\begin{equation}
\omega_{a}=\frac{\omega_s - M_2\sqrt{\omega_s^2-1+M_2^2}}{1-M_2^2}, \quad
k_{a}=\frac{\omega_s M_2 - \sqrt{\omega_s^2-1+M_2^2}}{1-M_2^2}.
\label{omegaaandka}
\end{equation}

It is readily seen that $k_a$ can be either positive or negative. The $k_a<0$ case represents acoustic waves traveling in the direction of the postshock flow, while the waves moving in the opposite direction towards the shock have $k_a>0$. The solution corresponding to shock oscillation frequency $\omega_s=1$ represents the standing acoustic waves that separate the solution traveling to the left $\omega_s>1$ from the one traveling to the right $(1-M_2^2)^{1/2}<\omega_s<1$ in the reference frame comoving with the postshock fluid. At large distances from the shock in the downstream region (far larger than the wavelength of the perturbations), the asymptotic pressure and the isentropic density perturbations are given by 
\begin{equation}
\bar{p}(x,y,\tau)=\bar{\rho}_a(x,\tau)= \mathcal{P}_s \cos\left(\omega_a \tau-k_a x \right)\cos(y),
\label{pa}
\end{equation}
where $\mathcal{P}_s$ is the amplitude of the pressure perturbations. The isentropic temperature variations induced by the acoustic radiation are just $\bar{T}_a(x,\tau) = \left(\gamma-1\right) \bar{p}(x,\tau)$.  

The acoustic contribution of the velocity perturbations are readily obtained through the Euler momentum equations
\begin{equation}
\frac{\partial \bar{u}_a}{\partial \tau}=-\frac{\partial \bar{p}}{\partial x} \quad \textrm{and}\quad \frac{\partial \bar{v}_a}{\partial \tau}=\bar{p}\ ,
\label{uvaeq}
\end{equation}
which can be used to write the long-time response
\begin{eqnarray}
\bar{u}_a(x,y,\tau)&=\mathcal{U}^a\cos\left(\omega_a\tau-k_a x \right)\cos(y), \label{ua}\\
\bar{v}_a(x,y,\tau)& =\mathcal{V}^a\sin\left(\omega_a\tau-k_ax \right)\sin(y),
\label{va}
\end{eqnarray}
where 
\begin{equation}
\mathcal{U}^a =\frac{ k_a}{\omega_a}\mathcal{P}_s \quad \textrm{and}\quad  \mathcal{V}^a =\frac{1}{\omega_a}\mathcal{P}_s
\label{UVa}
\end{equation}
are the associated amplitudes.

In absence of dissipative effects, vorticity disturbances downstream remain frozen to the fluid particles, with the amplitude being determined by the vorticity produced at the shock, namely
\begin{equation}
\bar{\omega}(x,y)=\frac{\partial \bar{v}}{\partial x}-\frac{\partial \bar{u}}{\partial y}=\left[\Omega_2\ \bar{p}_s\left(\tau=\frac{x}{M_2}\right) + \Omega_1  \cos\left(\frac{\omega_s}{M_2} x \right)\right]\sin(y)\,
\label{vort}
\end{equation}
where
\begin{equation}
\Omega_1=C_2\left[1+\left(\frac{k_x}{k_y}\right)^2\right]=C_2\left(1+\frac{1-M_2^2}{C_2^2 M_2^2}\zeta^2\right)
\label{Omega1}
\end{equation}
accounts for the shock-compression of the vortices, a one-dimensional effect, and
\begin{equation}
\Omega_2=\frac{M_2 \left(C_2-1\right) \sigma_a+ \sigma_b M_2 -1}{M_2}
\label{Omega2}
\end{equation}
indicates the contribution of shock deformations, which is a two-dimensional effect.

The linear perturbations in the velocity field satisfy the equation
\begin{equation}
\frac{\partial^2 \vec{\bar{v}}}{\partial \tau^2}= \nabla\times\nabla\times \vec{\bar{v}} + \nabla^2\vec{\bar{v}}\ ,
\label{uvec}
\end{equation}
with the breakdown of irrotational-acoustic and steady-rotational perturbations obeying, separately,
\begin{equation}
\frac{\partial^2 \vec{\bar{v}}_a}{\partial \tau^2}= \nabla^2\vec{\bar{v}}_a 
\quad \textrm{and}\quad
\nabla^2\vec{\bar{v}}_r = \frac{\partial \bar{\omega}}{\partial y}\hat{e}_x- \frac{\partial \bar{\omega}}{\partial x} \hat{e}_y \ .
\label{uvecar}
\end{equation}

With the acoustic field being given by~\eqref{ua} and \eqref{va}, the spatial distribution of the rotational-velocity perturbations is now derived to provide the complete velocity field. 
The solenoidal part is calculated by tracking the vorticity left behind by the oscillating shock front from $\tau=0$, yielding
\numparts
\begin{eqnarray}
\bar{u}_r(x,y) =& \left[ \bar{u}_p + \frac{\exp{\left(-x\right)}}{\sqrt{1-M_2^2}}\mathscr{P}_s \left(s=\frac{M_2}{\sqrt{1-M_2^2}}\right)\right]\cos{(y)}\ ,\\
\bar{v}_r(x,y) =& \left[ \frac{\partial\bar{u}_p}{\partial x}- \frac{\exp{\left(-x\right)}}{\sqrt{1-M_2^2}} \mathscr{P}_s\left(s=\frac{M_2}{\sqrt{1-M_2^2}}\right)\right]\sin{(y)}\ ,
\label{uvrot}
\end{eqnarray}
\endnumparts
with the particular solution $\bar{u}_p$ being provided by
\begin{eqnarray}
& \bar{u}_p (x)	= -\frac{2 \Omega_2}{\pi} \int_0^1 f(z) \frac{\cos{\left(z\ x \sqrt{M_2^{-2}-1} \right)}}{1+(M_2^{-2}-1)z^2}  {\rm d}z\ + 	\nonumber \\
&+\Omega_1  \frac{\cos{\left(\zeta\ x \sqrt{M_2^{-2}-1} \right)}}{1+(M_2^{-2}-1)\zeta^2}+\frac{2 \Omega_2 (1-C_2^{-1})}{\pi}\left(\zeta^2-\frac{C_2 M_2^2}{1-M_2^2}\right) \times  \label{up} \\
&\times\int_0^1  \frac{f(z)}{\zeta^2-z^2} \left[ \frac{\cos{\left(z\ x \sqrt{M_2^{-2}-1} \right)}}{1+(M_2^{-2}-1)z^2}  - \frac{\cos{\left(\zeta\ x \sqrt{M_2^{-2}-1} \right)}}{1+(M_2^{-2}-1)\zeta^2}\right]{\rm d}z, \nonumber
\end{eqnarray}
and with the auxiliary integration function $f(z)$ being defined in \eqref{fz1}.

The function \eqref{up} is qualitatively similar to that shown in \cite{Wouchuk2009} and \cite{Huete2017} for adiabatic and reacting shocks waves, respectively. Details of its derivation, omitted here for the sake of conciseness, involve the use of the inverse Laplace transform technique. The asymptotic rotational contribution of the velocity field is also written as a piecewise function of the shock oscillation frequency, with the longitudinal
\begin{equation}
\bar{u}_{r}(x\gg 1,y) =  \cos(y)\left \{ \begin{array}{ll}
\mathcal{U}_{lr}^r \cos\left( \frac{\omega_s}{M_2} x\right) + \mathcal{U}_{li}^r \sin\left( \frac{\omega_s}{M_2} x\right) & ,\zeta \leq 1 \\
\mathcal{U}_{s}^r \cos\left( \frac{\omega_s}{M_2} x\right) & ,\zeta \geq 1
	\end{array} \right.
\label{urasym}
\end{equation}
and transverse
\begin{equation}
\bar{v}_{r}(x\gg 1,y) =  \sin(y)\left \{ \begin{array}{ll}
\mathcal{V}_{lr}^r \sin\left( \frac{\omega_s}{M_2} x\right) + \mathcal{V}_{li}^r \cos\left( \frac{\omega_s}{M_2} x\right) & ,\zeta \leq 1 \\
\mathcal{V}_{s}^r \sin\left( \frac{\omega_s}{M_2} x\right) & ,\zeta \geq 1
	\end{array} \right.
\label{vrasym}
\end{equation}
contributions being characterized by their amplitudes
\begin{equation}
\mathcal{U}_{lr}^r=\frac{\Omega_2 \mathcal{P}_{lr}+\Omega_1}{1+(M_2^{-2}-1)\zeta^2},\quad 
\mathcal{V}_{lr}^r=\zeta\ \mathcal{U}_{lr}^r\sqrt{M_2^{-2}-1},\quad
\label{UrlriUs}
\end{equation}
and 
\begin{equation}
\mathcal{U}_{li}^r=\frac{\Omega_2 \mathcal{P}_{li}}{1+(M_2^{-2}-1)\zeta^2},\quad 
\mathcal{V}_{li}^r=-\zeta\ \mathcal{U}_{li}^r\sqrt{M_2^{-2}-1},
\label{Vrlri}
\end{equation}
for the streamwise and crosswise components in the long-wavelength regime, respectively, and
\begin{equation}
\mathcal{U}_s^r=\frac{\Omega_2 \mathcal{P}_s+\Omega_1}{1+(M_2^{-2}-1)\zeta^2}, \qquad \mathcal{V}_s^r=\zeta\ \mathcal{U}_s^r \sqrt{M_2^{-2}-1}
\label{UVs}
\end{equation}
for the corresponding short-wavelength regime amplitudes.

The density variations $\bar{\rho}_e$ due to the entropy waves is obtained from Rankine-Hugoniot relations~\eqref{massRH}-\eqref{eneRH} by subtracting the contribution of sonic waves:
\begin{equation}
\bar{\rho}_e(x,y)=\left(\mathcal{D}-1\right) \bar{p}_s\left(\tau=\frac{x}{M_2}\right)\cos(y),
\label{dene}
\end{equation}
where $\mathcal{D} = \left(2 M_2 \sigma_b -1\right)/M_2^2$ is the density perturbation amplitude relative to the pressure at the shock. The temperature variation corresponding to the entropy waves is given by $\bar{T}_e(x) = -\bar{\rho}_e(x)= -(\mathcal{D}-1) \bar{p}_s(\tau=x/M_2)$, where the temperature is normalized with the base flow temperature.

\begin{figure}
\centering
\includegraphics[width=0.99\textwidth]{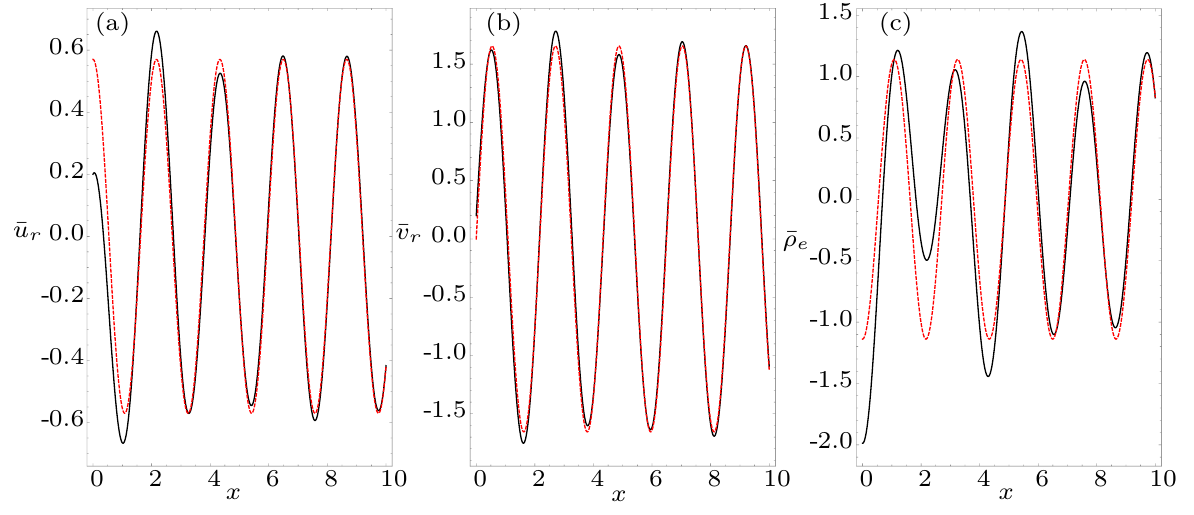}\\
\includegraphics[width=0.99\textwidth]{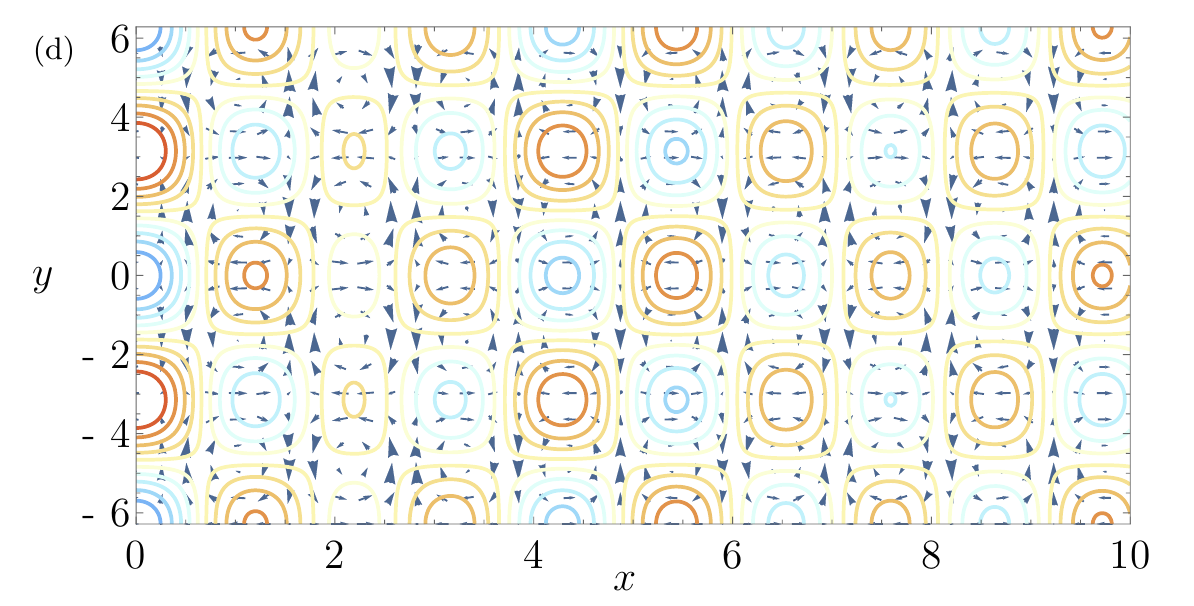}
\caption{Spatial distribution of the rotational velocity field and entropic density perturbations from the origin contact locus $x=0$. Panels (a) and (b) for the streamwise and transverse directions and panel (c) for the density function. Panel (d) shows iso-contours in the $(x,y)$ domain. Computations made for $M_1=2$, $\varepsilon=0.2$, and $\zeta = 1.5$.}
\label{fig:condenvort}
\end{figure}  

Figure \ref{fig:condenvort} shows the spatial distribution of the rotational velocity field and entropic density perturbations from the origin $x=0$ for $M_1=2$, $\varepsilon=0.2$ and $\zeta = 1.5$. The upper panels (a)-(c) display the functions $\bar{u}_r$, $\bar{v}_r$, and $\bar{\rho}_e$ as a function of $x$, respectively, and the lower panel (d) shows the vector velocity field superposed to the 
density iso-contours in the plane $(x,y)$. The exact solution for the rotational-velocity contribution (solid line) is found to approach the asymptotic solution (red-dashed line) in a fairly short distance, while the entropic-density function exhibits a longer transient period towards the long-time solution. 

The amplitude of the asymptotic rotational and acoustic velocity perturbations, as well as the corresponding entropic and acoustic density perturbations, have been conveniently expressed in terms of the shock pressure perturbation amplitude. Likewise, the asymptotic shock oscillation amplitude has been written as a function of $\mathcal{P}$ in equation \eqref{Jlri}. It is then immediate to obtain the value of any perturbation variable with the aid of Figure \ref{fig:J0}. Further computations, as the long-time amplitudes for the velocity and density perturbations, have been computed in Figures~A1-A3 of \cite{Huete2018}. As found in Figure~\ref{fig:J0}, there exists a peak in the perturbation amplitude near the critical frequency $\zeta=1$ and the effect of endothermicity, along with the associated amplitude change, is to stretch the peak in the frequency domain. This effect occurs for any perturbation variable: pressure, velocity or density. The acoustic contribution is found to provide a negligible contribution in comparison to the rotational or entropic counterpart.

\textcolor{black}{Density perturbations generated by the interaction of upstream asphericities with the shock are found to play a pivotal role in driving post-shock turbulence by buoyancy effects, which translates into a reduction of the critical neutrino luminosity necessary for producing explosion \cite{Mueller16,Mabanta18}. In \cite{Huete2018}, the effect of vorticity waves with isotropic distribution of orientations was considered. This is perhaps not an accurate representation. The convective motion is characterized by a dominant eddy with a specific size that undergo stretching due to the accelerated collapse, which increases the value of the shear angle $\theta$.}

\begin{figure}
\centering
\includegraphics[width=\textwidth]{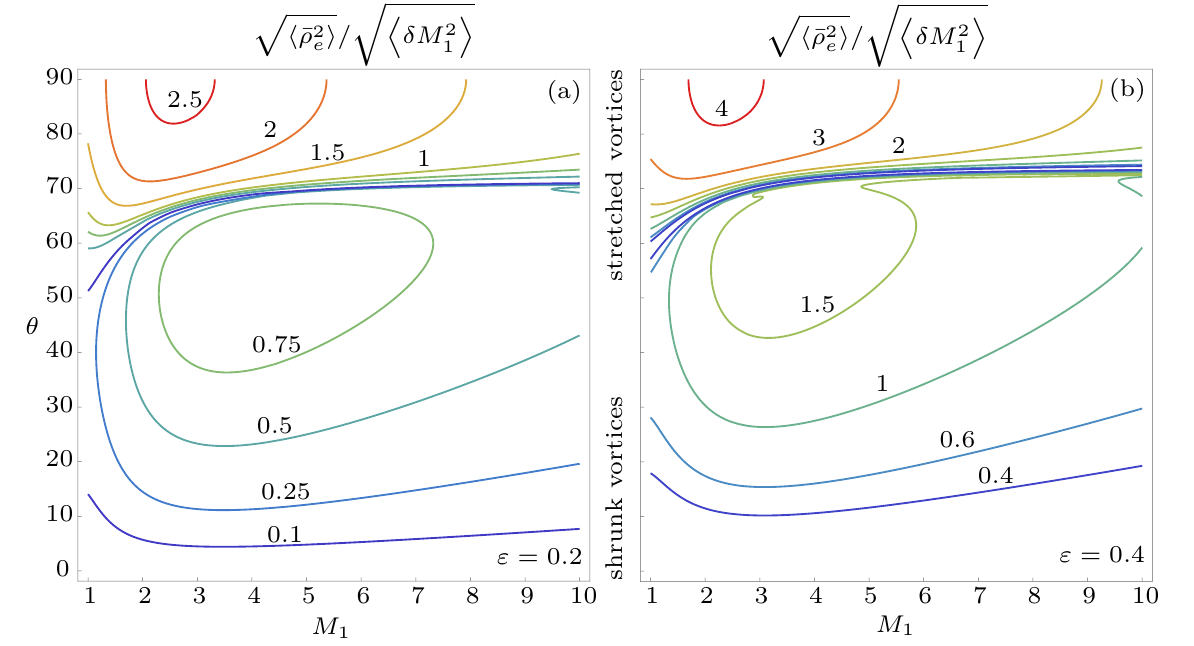}
\caption{Iso-contours of average entropic-density perturbations as a function of the shear-wave angle $\theta$ and the Mach number $M_1$, for $\varepsilon=0.2$ (a) and for $\varepsilon=0.4$ (b).}
\label{fig:deltaL}
\end{figure}

\textcolor{black}{Anticipating that the entropic contribution dominates the post-shock density perturbations, the correlation between the upstream turbulent Mach number and the entropic-density perturbations is used to evaluate the contribution of the shock-generated density fluctuations through the factor
\begin{equation}
\frac{\sqrt{\left< \bar{\rho}_e^2\right>}}{\sqrt{\left<\delta M_1^2\right>}} =\frac{M_2^2\mathcal{C}_2^2}{M_1}\frac{\left(\mathcal{D}-1\right)}{M_2^2\mathcal{C}_2^2+\zeta^2\left(1-M_2^2\right)}|\mathcal{P}|,
\end{equation}
which is computed in Figure~\ref{fig:deltaL} as a function of the upstream Mach number and the shear-wave angle $\tan \theta=k_y/k_x$. The upper region, corresponding to the highly-elongated vorticity limit $\zeta\ll1$ ($\theta\sim 90$), is found to yield stronger density perturbations than those produced for $\zeta\gg1$ ($\theta\sim 0$), the lower limit. Modification of the endothermic contribution $\varepsilon$ does not change the qualitative picture significantly, but it increases the figures. That is, despite pressure perturbations and shock-ripple deviations decrease with the endothermicity associated to the nuclear dissociation mechanism, entropic-density perturbations grow with the factor $\varepsilon$.}

\begin{figure}
\centering
\includegraphics[width=0.85\textwidth]{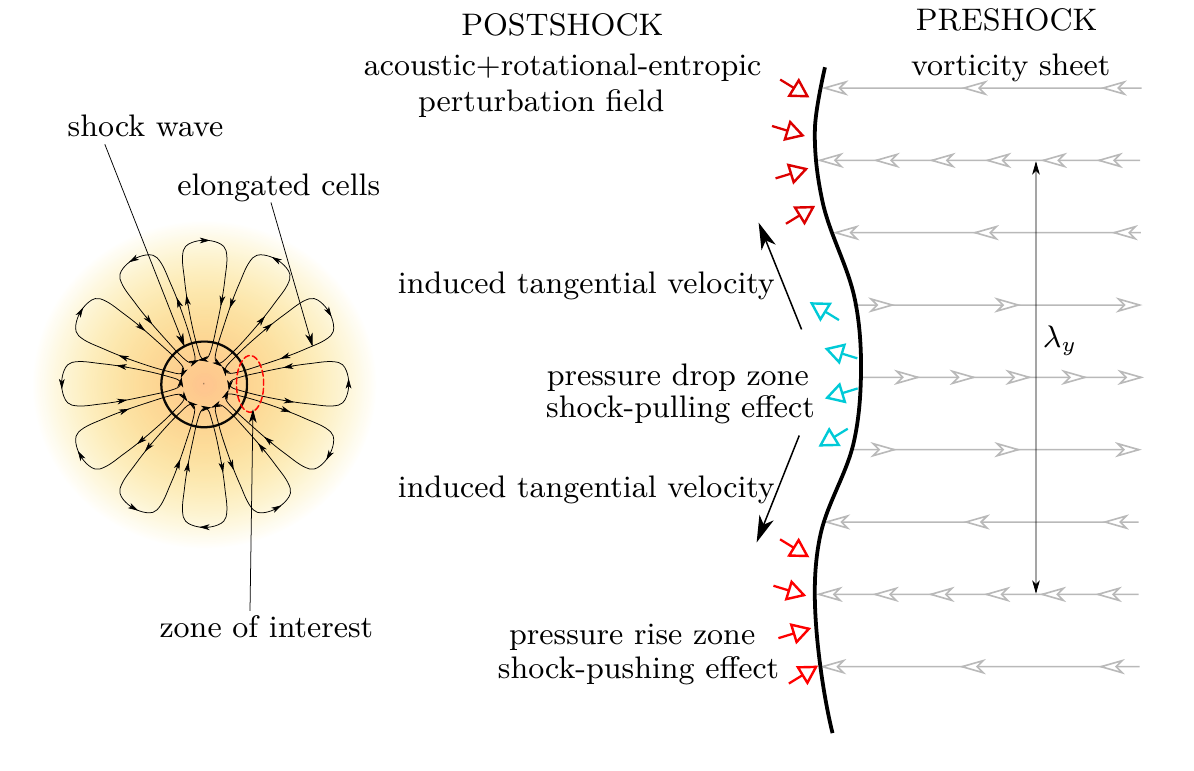}
\caption{Scheme of the interaction of the shock front with a vorticity sheet in the planar-shock limit, $R_{\rm shock}\gg\lambda_y$. As in Figure~\ref{fig:shock_vor}, induced transverse velocity behind the shock generates a counter-pressure effect to balance the shock rippling.}
\label{fig:shock_vor2}
\end{figure}

\textcolor{black}{
\section{Interaction with highly-stretched vortices}\label{S4}}
\textcolor{black}{
Natural convective cells are typically formed when temperature gradients are counter-aligned with the gravity field, conditions that are met in massive stars. When, in addition, they are advected by a highly-accelerating radially-converging flow, they elongate in the radial direction. In this scenario, the upstream perturbation field is strongly anisotropic, so that the canonical shock-turbulence interaction approach is no longer applicable. It is, however, possible to construct a representative problem setup with the formulation previously presented, as depicted in Figure \ref{fig:shock_vor2}.}

\textcolor{black}{
In the planar-shock limit, \textit{i.e.}, when the characteristic length of the upstream perturbation field is much smaller than the shock radius, the formulation of the problem shown in previous section can be particularized to the case $k_x\ll k_y$ or, equivalently, $\zeta\ll1$. In this slender limit, the acoustically-induced shock oscillation period is much shorter than the characteristic residence time of the convective cells crossing out the shock, thereby providing the following upstream modulation
\begin{equation}
\bar{u}_1 \left( x_1, y_1 \right)= \cos\left(k_y y_1 \right), \qquad {\rm and} \qquad  \bar{v}_1 \left( x_1, y_1 \right)=0, 
\end{equation}
as the transverse contribution becomes negligible.}

\textcolor{black}{
The corresponding boundary conditions at the shock front are appropriately adapted to yield
\begin{equation}
\frac{d \xi_s(r)}{d r} = \frac{\sigma_a}{\sqrt{1-M_2^2}} \bar{p}_s(r)+ \frac{1}{\sqrt{1-M_2^2}}
\label{xisrp2}
\end{equation}
and
\begin{equation}
\left.\frac{1}{r}\frac{\partial \bar{p}_s}{\partial \chi}\right|_s = -\sigma_b \frac{\partial \bar{p}_s(r)}{\partial r}-\frac{M_2^2 \left(C_2- 1\right)}{\sqrt{1-M_2^2}}\xi_s(r),
\label{psr2}
\end{equation}
which have been conveniently written as a function of the variables $r$ and $\chi$ \cite{Velikovich07}. After some straightforward manipulation, it is obtained the Laplace Transform of the pressure perturbations, namely
\begin{equation}
\mathscr{P}_s(s) =\frac{s^2 -\sigma^*}{s\left(s\sqrt{s^2+1}+\sigma_b s^2 + \sigma_c\right)},
    \label{PsLaplace2}
\end{equation}
provided that $\left(1+\sigma_b\right)\bar{p}_{s0}=1$, and with the factor in \eqref{sigma} being now
\begin{equation}
\sigma^*  =-\sigma(\zeta=0)=\frac{M_1^2\left(\kappa+1\right)-1}{\left(\gamma+1\right)\left(M_1^2-1\right)}.
    \label{sigmastar}
\end{equation}
}

\textcolor{black}{The initially planar shape of the shock front is distorted as a result of the interaction with the upstream modulated velocity field. For $y=0$, the shock encounters an always positive velocity perturbation aligned with the shock propagation, which pulls the shock front upwards. The opposite would apply for $k' y_1=y=\pi$, where an always negative velocity input pushes the front backwards. Then, the tangential velocity perturbation induced by the shock corrugation, with the associated mass flux, tends to restore the shock shape. This two counter-effects are responsible of the shock oscillations in the initial stage. Eventually, the shock approaches a steady-state regime when the two opposed effects balance, something that cannot occur when the upstream non-uniform flow is made of vortices with $k_x\sim k_y$. The corresponding long-time functions take the forms $\bar{p}_s\left(\tau\gg1,y\right)=\bar{p}_s^\infty\cos(y)$ and $\xi_s\left(\tau\gg1,y\right)=\xi_s^\infty \cos(y)$ for the shock pressure and ripple variables, respectively, with the associated amplitudes being
\begin{equation}
\bar{p}_s^\infty=-\frac{1}{\sigma_a}\qquad {\rm and} \qquad \xi_s^\infty=\frac{1}{\sigma_c\sqrt{1-M_2^2}}. 
\label{psxisasym2}
\end{equation}
}

\begin{figure}
\centering
\includegraphics[width=\textwidth]{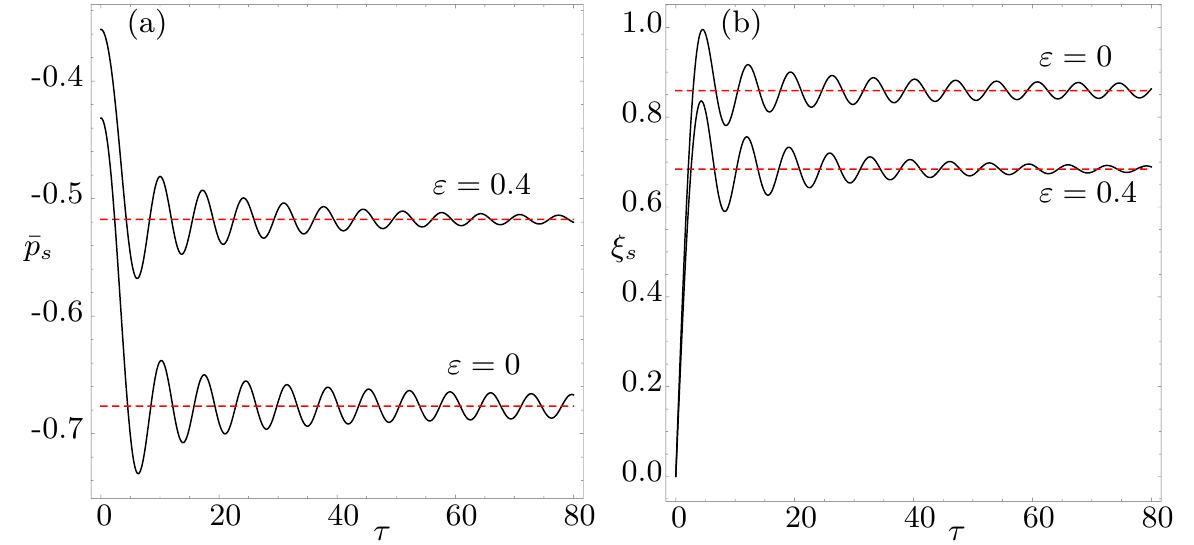}
\caption{Non-dimensional shock pressure $\bar{p}_s$ (a) and shock ripple $\xi_s$ (b) as a function of the dimensionless time $\tau$ for $M_1=5$, and for $\varepsilon=0.0$ and $\varepsilon=0.4$. Exact temporal evolution \eqref{xisr} in black-solid lines and asymptotic solution \eqref{psxisasym2} in red-dashed lines.}
\label{fig:psxis2}
\end{figure} 

\textcolor{black}{The evolution of the shock pressure and shock ripple perturbations, when the front travels through a stripe-like pattern of longitudinal perturbations, is computed in Figure~\ref{fig:psxis2} for $M_1=5$, and for $\varepsilon=0.0$ and $\varepsilon=0.4$. The transient behaviour is readily obtained by just taking the limit $\zeta-1\ll1$ in \eqref{pr} and \eqref{xisr}, respectively. It is observed that endothermic nuclear dissociation exhibits a stabilizing mechanism, since the long-time amplitudes are smaller and the oscillations decay faster for $\varepsilon=0.4$ than for $\varepsilon=0$. This is in consonance with Figure~\ref{fig:Bates}, where the case $M_1=5$ and $\varepsilon=0.4$ lies on the region $\eta<1$, while regular adiabatic shocks with $\varepsilon=0$ lie on the zone $\eta>1$.}

\textcolor{black}{In regard to the far-field perturbations, it is immediate to see that the acoustic contribution is negligible, as pressure perturbations decay exponentially with the distance from the shock for $\zeta<1$. What remains constant, in absence of diffusive effects, is the entropy-vorticity perturbation generated across the shock, namely
\begin{equation}
\bar{\rho}_e\left(x\gg1,y\right)=-\frac{\mathcal{D}-1}{\sigma_a}\cos(y),\quad   
\bar{\omega}\left(x\gg1,y\right)=\frac{\mathcal{C}_2\sigma_a-\Omega_2}{\sigma_a}\sin(y).
\label{rhoevort2}
\end{equation}
}

\textcolor{black}{Likewise, by direct inspection of \eqref{uvrot} and anticipating that tangential velocity generated behind the shock when $\bar{u}_r=$constant is of acoustic type, and then evanescent, it is found that 
\begin{equation}
\bar{u}_r\left(x\gg1,y\right)=\frac{\mathcal{C}_2\sigma_a-\Omega_2}{\sigma_a}\cos(y),\quad   
\bar{v}_r\left(x\gg1,y\right)=0, 
\label{rhoevr2}
\end{equation}
for the longitudinal and transverse components of the rotational velocity field.}

\section{Conclusions}\label{S5}
When a shock wave encounters a vorticity wave on its way, their interaction results in a deformation of the shock. The latter induces pressure changes that results in the radiation of acoustic waves downstream. Along with the sonic mode, post-shock perturbations include vortical and entropic disturbances that are convected by the fluid particles downstream. The distinguished feature of nuclear-dissociating shocks is the endothermic contribution, which depends on the shock intensity relative to the flow stream. Then, in a likely non-uniform context, perturbations ahead of the shock as the shear-pattern considered in this work, the amount of nuclei that are dissociated is affected by the shock perturbation, which in turn affects the energy balance across the shock, and ultimately the amplitude of the perturbations downstream. This effect can be easily studied when considering the interaction of the shock with intermediate vortical scales, \textit{i.e.}, those whose characteristic length is sufficiently small for the shock to be considered a planar front, yet sufficiently large for the shock to be a seen as a discontinuity front. 

This problem is particularly relevant in the context of core-collapse supernova explosions, where the shock after bounce swallows the convective structures generated upstream within the supersonic inwards-traveling mass. The shock-perturbation interaction modifies and creates additional perturbations downstream, which affects the critical conditions for the supernova explosion. \textcolor{black}{When the upstream flow is assumed to be dominantly isotropic, theory predicts that the injection of non-radial motion and the buoyancy-driven convection triggered by entropy waves reduce the critical neutrino luminosity by$\sim$ 12-24 per cent, for typical problem parameters \cite{Mueller16, Abdikamalov16, Huete2018}. When convective cells are not sufficiently turbulized by the inwards acceleration of the fluid particles, the isotropic assumption may be inaccurate. Motivated by this fact, in this work the emphasis is placed in bidimensional structures made of iso-density vortices. The temporal evolution of the oscillating shock and the long-time asymptotic expressions have been derived analytically as a function of the dominant governing parameters: the shock strength $M_1$, the nuclear dissociation degree $\varepsilon$, and the incident shear-wave angle $\sim k_x/k_y$. Likewise, the exact and asymptotic spatial distribution of the perturbations in the shocked gas have been derived explicitly with use made of the Laplace Transform technique. The limit of high elongated vortices, corresponding to $k_x\ll k_y$, is also evaluated in terms of closed-form expressions.}

The effect of the endothermic nuclear dissociation is found to diminish the amplitude of the shock oscillations and to reduce the acoustic radiation. Contrarily, the entropic density disturbances grow with the factor $\varepsilon$.
In some conditions, for sufficiently endothermic shock, the transient evolution is significantly shorten. The distribution of the downstream entropic and vortical perturbations have been provided analytically. These perturbations serve as initial conditions of the post-shock regime, as density variances become buoyant in a temporal scale that exceeds the shock influence. This phenomenon will be studied in more depth in a future work. 

\section*{Acknowledgements}
This work is  supported by the Ministry of Science, MEC (ENE2015-65852-C2-1-R) and Fundaci\'on Iberdrola Espa\~na (BINV-hBbhOeJQ), Spain (for C. Huete), by MES RK state-targeted program BR05236454, MES RK grant No. 346, NU ORAU grant SST 2015021 and NU grant No. 090118FD5348 (for E. Abdikamalov).

\end{document}